\documentclass[usenatbib]{mn2e}
\usepackage{graphicx}
\begin{document}

\author[A.L.Watts et al]{A.L.Watts$^1$, N.Andersson$^1$, 
H.Beyer$^2$ and B.F.Schutz$^2$
\\ $^1$ Department of Mathematics, University of Southampton, Southampton SO17 1BJ, UK \\
$^2$ Max-Planck Institut f\"ur Gravitationsphysik, 
Albert Einstein Institut f\"ur Gravitationsphysik, Am M\"uhlenberg 1, 14476 Golm, Germany}

\title[Differentially rotating spherical shells: Normal modes]{The
oscillation and stability of differentially rotating spherical shells: The normal-mode
problem}

\maketitle

\begin{abstract}
An understanding of the dynamics of differentially rotating systems is key
to many areas of astrophysics.  We investigate the oscillations of a simple system 
exhibiting differential rotation, and discuss issues concerning 
the role of corotation points and the emergence of 
dynamical instabilities.  This problem is of particular relevance to the
emission of gravitational waves from oscillating neutron stars, which are
expected to possess significant differential rotation immediately after
birth or binary merger.  
\end{abstract}


\section{Background and motivation}

This paper describes a dynamical problem that is relevant to many areas of
astrophysics. We consider the oscillations of a simple system 
exhibiting differential rotation, and discuss issues concerning 
the role of corotation points and the emergence of 
dynamical instabilities. This is known to to be a technically challenging 
problem (analogous to that of Couette flow in standard fluid dynamics)
that remains to be understood in detail. 
In addition to being mathematically interesting, the problem is of potentially great 
astrophysical relevance since
the oscillations of compact objects such as neutron stars or strange stars are
a promising source of detectable gravitational waves.   Of particular
interest are oscillations that can grow via either a dynamical
instability or a secular instability to the emission of gravitational waves, such as the
Chandrasekhar-Friedman Schutz (CFS) instability \citep{fri78a,fri78b}. 
The most promising instability scenarios concern hot newly born 
neutron stars, which one would expect to possess some differential
rotation.  This expectation is borne out by the latest studies 
of rotational core collapse, which indicate that the remnant will indeed emerge from the
collapse rotating differentially \citep{zwe97, dm02a}.
Subsequent accretion of supernova remnant material \citep{wat02}, or
material from 
a companion \citep{fuj93}, may also drive differential rotation.
Differential rotation may also be maintained by the oscillations
themselves.  Recent work on r-modes in neutron stars suggests that the
modes can drive a uniformly rotating star into differential rotation via
non-linear effects \citep{rez00,lev01,lin01}. A differentially rotating massive neutron star may
also be generated following a binary neutron star merger. The maximum mass
of a differentially rotating merger remnant can exceed significantly the
maximum mass of a uniformly rotating remnant, remaining stable
over many dynamical timescales \citep{shi00a, bau00}.

It is clear that differential rotation is
an important factor in modelling these systems. Nevertheless, the vast 
majority of secular instability studies to date
have focused on uniformly rotating neutron stars (exceptions include
studies by \citet{ima95}, \citet{yos02}, and \citet{kar02}).  The situation
is somewhat  
different for the dynamical bar mode instability (see for example
\citet{shi00} and \citet{new01}). There have been a number
of large scale numerical simulations aimed at understanding the bar mode 
instability. These studies
provide significant insights into the  
dynamical stability of differentially rotating systems. 
Yet it is clear that a better theoretical understanding of 
this problem is desirable. This need is well illustrated by the 
recent discovery of dynamical instabilities that become 
active at very low values of $T/|W|$ (the ratio of rotational kinetic energy to gravitational 
potential energy) \citep{cen01,shi02}. 

There are two issues associated with differential rotation that make this
problem particularly interesting.  Firstly, differential rotation may
introduce new dynamical instabilities \citep{bal85,luy90a}.  Secondly, the
dynamical equations are complicated by the presence of corotation points.
These are points where the pattern speed of a mode matches the local
angular velocity, and at these points the governing equations are formally
singular.  This gives rise to a continuous spectrum and could in principle
permit other corotating solutions.  These two important issues have however
attracted little attention in the literature.
In particular, it is not at all clear what the dynamical role of the continuous
spectrum may be.

\citet{kar01} studied r-mode oscillations in rapidly and
differentially rotating Newtonian polytropes for two particular rotation
laws.  They found that the r-mode frequencies shifted towards the 
corotation region as the amount of differential rotation
increased and at high differential rotation the codes failed, apparently due to the
presence of corotation points. Solutions with corotation points were not,
however,
investigated.  Two subsequent studies of a differentially rotating
 shell for the same two rotation laws also focused on modes
without corotation points \citep{rez01, abr02}.  The authors claimed that
for these rotation laws the lowest frequency r-modes did
not develop corotation points, even at high differential rotation. 

For papers that deal more fully with dynamical instabilities and corotation
points we must go back a number of years (or consider the related problem of
differentially rotating disks for which there is a vast literature). 
In the 1980's Balbinski carried
out a number of studies on $z$-independent modes of differentially rotating
cylinders (for cylindrical coordinates $\varpi, \phi, z$) that addressed
both issues \citep{bal84a, bal84b, bal85}.  The first paper reported an
analytic solution for the continuous spectrum for one particular rotation
law.  Balbinski then solved the initial value problem for simple initial
data, and found that the physical perturbation associated with the continuous
spectrum died away on a dynamical timescale.   The second
paper examined the secular stability of the continuous spectrum 
to the emission of gravitational waves.  The final paper investigated the
behaviour of modes
that had no corotation points at low differential rotation.  As the degree
of differential rotation increased, some of the modes were seen to enter
corotation and in due course merge to give rise to a dynamical
instability.
In the early 1990's, Luyten generalised the work in \citet{bal85} to other
rotation laws \citep{luy90a}.  In two subsequent papers he solved for the
same $z$-independent modes in self-gravitating homogeneous masses rather than on
cylinders \citep{luy90b,
luy91}. Although Luyten's method could not
track modes with corotation points, modes appeared to enter and
leave the corotation region as the degree of differential
rotation increased.  Luyten too reported finding an instability at high
differential rotation.

There is clearly a need to analyse modes with a
$z$-dependence such as the toroidal r-modes. As a first step towards this goal, 
we present
a detailed analysis of the modes of a differentially rotating spherical shell.
We hope to improve our understanding of the technical challenge posed
by the presence of corotation points, and develop techniques 
that may later be generalised to the full problem of differentially 
rotating stars.  By investigating dynamical instabilities we may also be
able to rule out certain rotation laws, or set a threshold for the maximum
degree of differential rotation that could be sustained. 
We select four rotation laws from the literature in order to compare our
results to previous work, and one new law.  We note however that in all but one case these
laws were selected for mathematical convenience 
rather than
physical reasons. One would clearly prefer
to use a rotation law for a young neutron star that is grounded in
physics, emerging for example from simulations of core collapse
\citep{zwe97,dm02a}.  We hope to do this in future work.  

\section{Formulating the problem}

We consider an incompressible fluid confined to move in a thin spherical
shell.  We take the shell radius to be $R$.  Its thickness is taken to be
so small that we do not have to account for dynamics in the radial direction.  On these
assumptions, and taking the perturbation to depend on exp($im\varphi$), the
 $\theta$-component of the Euler equation simplifies to
\begin{equation}
\label{thetaeuler}
\partial_t \delta v_\theta + im\Omega \delta v_\theta -
2\Omega\cos\theta\delta v_\varphi = -\frac{1}{\rho R} \partial_\theta
\delta P, 
\end{equation}
and the $\varphi$-component to
\begin{equation}
\label{phieuler}
\partial_t \delta v_\varphi + im\Omega \delta v_\varphi + \tilde{\Omega} \delta
v_\theta = -\frac{im}{\rho R \sin\theta} \delta P, 
\end{equation}
where the equilibrium vorticity is
\begin{equation}
\tilde{\Omega} = 2\Omega \cos\theta + \sin\theta \partial_\theta \Omega.
\end{equation}
The continuity equation for an incompressible fluid is 
\begin{equation}
\partial_\theta (\sin\theta \delta v_\theta) + im\delta v_\varphi = 0.
\end{equation}
This tells us that only toroidal modes are permitted in this problem.  
Defining the stream function (in the standard way) as
\begin{equation}
U = \delta v_\theta \sin\theta, 
\end{equation}
we can combine equations (\ref{thetaeuler}) and (\ref{phieuler}) to give
the vorticity equation
\begin{equation}
\label{vorthetat}
(i\partial_t - m\Omega) \nabla_\theta^2 U + \frac{m\partial_\theta
\tilde{\Omega}}{\sin\theta} U = 0, 
\end{equation}
where 
\begin{equation}
\nabla^2_\theta = \partial_\theta^2 +
\frac{\cos\theta}{\sin\theta}\partial_\theta - \frac{m^2}{\sin^2\theta}
\end{equation}
is the Laplacian on the unit sphere. Changing variables to $x = \cos\theta$ the
vorticity equation becomes, 
\begin{equation}
\label{voryt}
(i\partial_t - m\Omega) \left[ (1-x^2)\partial_x^2 - 2x\partial_x -
\frac{m^2}{1-x^2} \right] U - m \partial_x \tilde{\Omega} U = 0.
\end{equation}
In terms of $x$ the equilibrium vorticity is 
\begin{equation}
\tilde{\Omega} = 2x\Omega - (1-x^2) \partial_x\Omega.
\end{equation}
Equation (\ref{voryt}) is singular at the poles ($x=\pm 1$).  Performing a
Frobenius expansion around these points,
\begin{equation}
U = (1\pm x)^\alpha \sum_{k=0}^\infty a_k(1\pm x)^k
\end{equation}
we find that the regular solution has $\alpha = m/2$.  Factoring out this
behaviour and defining a new variable $u$ by
\begin{equation}
U = u(1-x^2)^{m/2}, 
\end{equation}
we find that $u$ obeys the  equation
\begin{eqnarray}
\label{vorut}
(m\Omega - i\partial_t) \left[ (1-x^2)\partial_x^2 - 2(m+1) x\partial_x -
m(m+1)\right] u {} \nonumber \\ + m \partial_x \tilde{\Omega} u = 0.
\end{eqnarray}
We will use both equations (\ref{voryt}) and (\ref{vorut}) in this paper.

\section{Uniform rotation: The r-modes}
\label{uni}

Before we proceed to discuss differential rotation it is useful to 
recall the standard result for uniform rotation. After all, 
we are interested in understanding how the mode spectrum changes
as differential rotation is introduced.  The oscillations of a
uniformly rotating incompressible shell have been studied by a number of authors
\citep{hau40, ste69, lev01}, and are sometimes referred to
as Rossby-Haurwitz waves. Rossby-Haurwitz waves are however the limit on
the shell of the r-modes of stellar perturbation theory, and we will use
the latter term.  For uniform rotation we have
\begin{equation}
\partial_\theta\tilde{\Omega} = -2\sin\theta\Omega.
\end{equation}
Assuming a mode solution with time dependence $\propto$ exp($-i\sigma t$)
equation (\ref{vorthetat}) becomes
\begin{equation}
\label{uniformv}
(\sigma - m\Omega) \nabla_\theta^2 U - 2m\Omega U = 0.
\end{equation}
Assuming that $\sigma \ne m\Omega$ we have 
\begin{equation}
\nabla_\theta^2 U = \frac{2m\Omega}{\sigma - m\Omega} U.
\end{equation}
Expanding $U$ in spherical harmonics 
\begin{equation}
U = \sum_l C_l^m Y_l^m, 
\end{equation}
then by Legendre's equation we must have
\begin{equation}
\nabla_\theta^2 U = - \sum_l l(l+1) C_l^m Y_l^m.
\end{equation}
Thus we must have
\begin{equation}
- \sum_l l(l+1) C_l^m Y_l^m = \frac{2m\Omega}{\sigma - m\Omega} \sum_l
  C_l^m Y_l^m, 
\end{equation}
from which we deduce that the mode solutions are such that  
(i) only a single $C_l^m$ is non-vanishing and
(ii) the corresponding frequency is 
\begin{equation}
\sigma = m\Omega - \frac{2m\Omega}{l(l+1)}.
\end{equation}
These are the familiar r-modes.  

There is, however, another possible solution, the geostrophic solution.  Consider
what happens if we set $\sigma = m\Omega$.   It is clear from equation
(\ref{uniformv}) that the only possible solution in this case is $U = 0$,
which corresponds to a zero velocity perturbation.  Consider however what
this means for the displacement $\vec{\xi}$. In the case of uniform rotation, 
\begin{equation}
\delta v_\theta  = i(m\Omega - \sigma) \xi_\theta, 
\end{equation}
and
\begin{equation}
\delta v_\varphi  = i(m\Omega - \sigma) \xi_\varphi.
\end{equation}
It is clear that, if $\sigma = m\Omega$, a vanishing velocity perturbation is compatible with a
non-zero displacement.  This solution, trivial in the case of uniform
rotation, must be considered when the rotation
is differential as it is this solution that gives rise to the continuous spectrum.
Note the similarity with the standard r-modes, which are a 
trivial solution of the perturbation equations for a non-rotating
star.  

\section{The normal mode problem}

Let us now consider the case of a differentially rotating shell.
Assuming a mode solution with time dependence $\propto$ exp($-i\sigma t$)
equation (\ref{vorut}) becomes
\begin{eqnarray}
\label{umode}
(m\Omega - \sigma) \left[ (1-x^2) u'' - 2(m+1) x u' -
m(m+1)u\right] {}\nonumber \\ + m \tilde{\Omega}' u = 0, 
\end{eqnarray}
where the primes indicate derivatives with respect to $x$. At the poles, we require that the solution be
regular.  In addition, symmetry allows us to require that
$u$ be either even or odd under reflection through the equator.  This
implies that we can set boundary conditions on the equator that either $u$
or $u'$ be zero there.  These boundary conditions
allow us to solve the equation on the range $0\le x \le 1$ rather than
on the full range $-1 \le x \le 1$.   

If $\sigma \ne m\Omega(x)$ for all points on the shell we have a regular eigenvalue problem.  
If, however, $\sigma = m\Omega(x)$ at a point $x = x_c$, then $x_c$ is a corotation
point and equation (\ref{umode}) is formally singular. 
Corotation points are points where the 
pattern speed of a perturbation is equal to the local angular
velocity. Note that the term corotation point is a little misleading as
$x=x_c$ really represents a line of latitude on the shell. Equation
(\ref{umode})  has three classes of solution:
\begin{itemize}
\item Discrete real-frequency modes outside corotation, i.e. for which $\sigma\ne m\Omega(x)$ for
$|x| \le 1$.
\item Discrete complex frequency modes for which $\sigma = \sigma_r +
i\sigma_i$.   Dynamically unstable (growing) modes correspond to $\sigma_i
> 0$.
\item Corotating solutions for which  $\sigma = m\Omega(x)$ at
a corotation point $x_c$,  where $|x_c| \le 1$.
\end{itemize}
The first two kinds of solutions are easy to determine numerically 
for any given rotation law. In contrast, the solutions with corotation points
present a real challenge. Nonetheless, these solutions are of great
interest, not least because of their association with the continuous spectrum.
For this reason much of the following discussion will concern solutions of the third kind.
In the following sections we shall refer to solutions with corotation points as
being ``in corotation''.  Modes without corotation points that subsequently
develop them as the amount of differential rotation is increased will be
referred to as ``crossing the corotation boundary''.  It should be noted,
however, that solutions in corotation are in fact only corotating at one
particular latitude.  

\section{A necessary stability criterion}
\label{dyncon}

Before discussing detailed solutions for particular rotation laws it is 
useful to check whether unstable mode solutions are likely to exist.
Such solutions would typically correspond to dynamical instabilities, 
and hence they are of great physical significance.  

For inviscid parallel shear flow it is possible to derive a necessary condition for
stability that depends on the form of the rotation law, the Rayleigh
inflexion point criterion (see for example \citet{dra82}).
Here we derive a similar condition for modes of a
differentially rotating shell.  We start with equation
(\ref{voryt}), assuming a mode solution with complex $\sigma$,
\begin{equation}
\label{ymode}
(m\Omega -\sigma) \left[ \frac{d}{dx}[(1-x^2)U'] -
\frac{m^2 U}{1-x^2} \right] + m \tilde{\Omega}' U = 0.
\end{equation}
Multiply by the complex conjugate $U^*$ and integrate to get 

\begin{eqnarray}
\label{ins1}
\left[(1-x^2)U'U^*\right]_0^1 - \int_0^1 (1-x^2)|U'|^2 dx & - & {}\nonumber \\ m^2 \int_0^1 \frac{|U^2|}{1-x^2} 
dx  + m\int_0^1 \frac{\tilde{\Omega}'|U^2|}{m\Omega - \sigma}dx & = & 0.
\end{eqnarray}
The first term is clearly zero due to the boundary conditions. The 
imaginary part of equation (\ref{ins1}) is
\begin{equation}
\label{ins2}
m \sigma_i \int_0^1 \frac{\tilde{\Omega}' |U|^2}{|m\Omega - \sigma|^2}dx
= 0.
\end{equation}
For dynamically unstable modes $\sigma_i \ne 0$ and equation (\ref{ins2}) can
only hold if $\tilde{\Omega}'$ changes sign at some point on the interval.  This
is a necessary (although not sufficient) condition for instability.

\section{Corotating solutions}
\label{discorot}

We begin our study of corotating solutions  by considering the case
when there is a corotation point in the domain 
 $0<x_c<1$ (the symmetry of the problem means that there will be
a second corotation point in the lower hemisphere, at $x=-x_c$).  We
will consider the special cases of $x_c = 1, 0$ (corotation point at
the poles or the equator) later. Employing the method 
 of Frobenius to
investigate the solutions in the vicinity of the corotation point, we let
\begin{equation}
u = \sum_{n=0}^\infty a_n (x-x_c)^{n+\gamma}.
\end{equation}
For all of the rotation laws examined in this paper $m\Omega - \sigma$
contains only a single factor of $x-x_c$, also $\Omega$ is an analytic
function of $x$.  Assume also (for now) that $\tilde{\Omega}'$ does not have a zero at $x=x_c$.
Substituting into equation (\ref{umode}) to obtain the indicial
equation, we find that $\gamma = 0,1$, so we will have one regular and one singular
solution.  The general solution is 
\begin{equation}
\label{gensol}
u = u_{reg} + u_{sing}, 
\end{equation}
where
\begin{equation}
\label{ureg}
u_{reg}  = \sum_{n=0}^\infty a_n (x-x_c)^{n+1}, 
\end{equation}
and
\begin{equation}
\label{using}
u_{sing}  =  \sum_{n=0}^\infty \left[a_n (x-x_c)^{n+1}\ln(x-x_c)+ c_n (x-x_c)^{n}\right].
\end{equation}
Note that although $u_{sing}$ is finite and continuous at $x_c$ its
derivatives are singular.

\subsection{Possible regular solutions}
\label{sreg}

Let us begin by asking whether the problem admits any solutions that
are regular at corotation (i.e. contain $u_{reg}$ only).  First of all note that if $u$ is
regular in corotation then so is $U$.  We 
now apply the method that was used to derive a necessary condition for
instability in section \ref{dyncon}.  

We begin with equation (\ref{ymode}),
multiply by $U^*$ and as before integrate.  This time, however, the frequency 
is taken to be real and we are going
to treat the two domains $0\le x \le x_c$ and $x_c \le x \le 1$
separately.  On the first domain
\begin{eqnarray}
\left[(1-x^2)U'U^*\right]_0^{x_c} - \int_0^{x_c} (1-x^2)|U'|^2 dx & - & {}
\nonumber \\ m^2 \int_0^{x_c} \frac{|U^2|}{1-x^2} 
dx+ m\int_0^{x_c} \frac{\tilde{\Omega}'|U^2|}{m\Omega - \sigma}dx & = & 0.
\end{eqnarray}
The first term vanishes due to boundary conditions and we are left with
\begin{eqnarray}
-  \int_0^{x_c} \left[(1-x^2)|U'|^2 +  \frac{m^2|U^2|}{1-x^2} \right]
dx & + & {} \nonumber \\ m\int_0^{x_c} \frac{\tilde{\Omega}'|U^2|}{m\Omega
- \sigma}dx & = & 0.
\end{eqnarray}
It is clear that we must have
\begin{equation}
m\int_0^{x_c} \frac{\tilde{\Omega}'|U^2|}{m\Omega - \sigma}dx > 0.
\end{equation}
(Note that  $U/( m\Omega - \sigma)$
is well behaved at $x_c$ if $u \propto u_{reg}$.)
Treating the second domain in the same way we can derive a
second condition, which is that we must have
\begin{equation}
m\int_{x_c}^1 \frac{\tilde{\Omega}'|U^2|}{m\Omega - \sigma}dx > 0.
\end{equation}

Now $m\Omega - \sigma$ clearly changes sign at the corotation point.  So
unless $\tilde{\Omega}'$ also changes sign somewhere on the shell it will be
impossible to satisfy both conditions.  A rotation law will therefore only admit
purely regular solutions if $\tilde{\Omega}'=0$ somewhere on the shell.

There is one other class of regular solution that might be
encountered. The governing equation (\ref{umode}) will remain non-singular
if, for a particular frequency, the zero of $m\Omega - \sigma$
coincides with that of $\tilde{\Omega}'$.  This will in fact always be the
case for some frequency in the continuous spectrum range if
$\tilde{\Omega}'$ has a zero on the shell.  Unlike the regular
Frobenius solution $u_{reg}$, however,  solutions in this class need not
have a zero in the eigenfunction at the corotation point. 

\subsection{General (singular) solutions}
\label{singsol}

The coefficients of both the regular and singular solutions (equations
(\ref{ureg}) and (\ref{using})) are found by substituting the
solutions into equation (\ref{umode}).   We find $c_0$ to be related to
$a_0$, 

\begin{equation}
\label{a0c0}
c_0 = \frac{\Omega'(x_c)}{\tilde{\Omega}'(x_c)}(x_c^2 - 1) a_0.
\end{equation}
Hence $a_0$ and $c_1$ (equation \ref{using}) are the two arbitrary constants in the
general power series solution.

The Frobenius expansion gives us an approximation to the general solution
 close to the corotation point. This solution is useful in a numerical search for
solutions.  However, the logarithm term
causes a problem. In particular, it  gives rise to a branch point at $x_c$ 
and we would in principle
need to integrate around this point in the complex plane.  An alternative is to seek an
equivalent well-behaved solution for $x<x_c$ and match the two solutions at
the corotation point.  

Following \citet{bal85} we make a change of variable
\begin{equation}
z = \frac{x_c}{x}.
\end{equation}
The corotation point is now at $z = 1$.  Equation (\ref{umode}) becomes 
\begin{eqnarray}
(m\Omega - \sigma) \Big[\frac{(z^2 - x_c^2)}{x_c^2} (z^2 u'' + 2zu') +
2(m+1)zu'  -  {} \nonumber \\ m(m+1)u\Big]  + mu\tilde{\Omega}' =  0.
\end{eqnarray}
As before, we look for a power series solution in the vicinity of the
corotation point, letting
\begin{equation}
u = \sum_{n=0}^\infty b_n (z-1)^{n+\gamma}.
\end{equation}
Once again the indicial equation yields $\gamma = 0,1$, and the general solution is 
$u = u_{reg} + u_{sing}$, where
\begin{equation}
\label{uzreg}
u_{reg}  = \sum_{n=0}^\infty b_n (z-1)^{n+1},
\end{equation}
and
\begin{equation}
\label{uzsing}
u_{sing}  =  \sum_{n=0}^\infty \left[b_n (z-1)^{n+1}\ln(z-1)+ d_n (z-1)^{n}\right].
\end{equation}
The free parameters in the general solution are $b_0$ and $d_1$.
In the region where $x<x_c$, $z > 1$.  So by writing the
equation in terms of $z$ we have found a singular solution whose
logarithmic terms are real in the region where $x<x_c$.  

When solving for corotating solutions we follow \citet{bal85} and use the two
different solutions in the regions where
the logarithms have positive argument, and match at the corotation point.  Denoting
by $u_R$ the solution in the region where $x>x_c$, and by $u_L$ the
solution in the region where $x<x_c$ (and $z>1$), we have

\begin{eqnarray}
u_R & \approx & c_0  +  (a_0 + c_1)(x-x_c)  {}\nonumber \\ &&  + [a_0 + a_1(x-x_c)
](x-x_c)\ln|x-x_c|, \\
u_L & \approx & d_0  +  (b_0 + d_1)(z-1) {} \nonumber \\ && + [b_0 + b_1(z-1)
 ] (z-1)\ln|z-1| \nonumber \\
& \approx & d_0  +  (b_0 + d_1)\frac{(x_c - x)}{x} {}\nonumber \\ && + \left[b_0 + b_1\frac{(x_c - x)}{x} 
\right] \frac{(x_c - x)}{x}\ln\left|\frac{x_c - x}{x}\right|.
\end{eqnarray}
In the limit as the
solutions approach the corotation point, all of the terms tend to zero
apart from the constant terms.  For $u$ to be continuous we must have
\begin{equation}   
c_0 = d_0.
\label{c0d0}
\end{equation}
Now consider the first derivatives.  

\begin{eqnarray}
\frac{du_R}{dx} & \approx & 2a_0 + c_1 +  (3a_1 + 2c_2)(x-x_c) {} \nonumber
\\ && +
a_0\ln|x-x_c| +  2a_1(x-x_c) \ln|x-x_c|,  \\
\frac{du_L}{dx} & = & - \frac{z^2}{x_c} \frac{du_R}{dz}  \nonumber {}
\\ & \approx & - \frac{x_c}{x^2}\Big[2b_0 + d_1 +  (3b_1 +
2d_2)\frac{(x_c-x)}{x}{} \nonumber
\\ && +
b_0 \ln\left|\frac{x_c-x}{x}\right| +
2b_1 \frac{(x_c-x)}{x} \ln\left|\frac{x_c-x}{x}\right| \Big].
\end{eqnarray}
In the limit as the derivatives approach the corotation point all of the
terms tend to zero apart from the constant terms and the plain logarithmic
terms. We get
\begin{eqnarray}
\lim_{x \rightarrow x_c^+} \frac{du_R}{dx} & = & 2a_0 + c_1 + a_0 \lim_{x \rightarrow x_c} \ln
|x - x_c|,  {} \\
\lim_{x \rightarrow x_c^-} \frac{du_L}{dx}&  = & -\frac{1}{x_c}\Big[ 2b_0 + d_1
 + b_0 \lim_{x \rightarrow x_c} \ln
|x - x_c|{}\nonumber \\ && - b_0 \ln x_c\Big].
\end{eqnarray}
Our insistence on the continuity of the
function (equation (\ref{c0d0})) means that 
\begin{equation}
b_0 = -x_c a_0.
\label{cond1}
\end{equation}
Recall that $a_0$ and $c_0$ are related by equation (\ref{a0c0}),
and so the coefficients of the singular $\ln|x-x_c|$ term are equal at the corotation
point.  But in addition to this logarithmic term there is also a finite
step in the first derivative.  By permitting a step in the first
derivative there will  be a solution that meets the boundary
conditions for any frequency within the range of the continuous
spectrum.  In fact, for each
frequency there will be two solutions, one of each symmetry, with a
particular step in the first derivative, that meet the boundary
conditions (note that for a given frequency the step size for the solutions of
opposite symmetry need not be the same). 

The step in the first derivative varies across the range of continuous
spectrum frequencies and it can, for certain frequencies, be zero.
If the step in the first derivative vanishes then
\begin{equation}
2a_0 + c_1 = -\frac{1}{x_c}(2b_0 + d_1 - b_0 \ln x_c).
\label{cond2}
\end{equation}
Using equation (\ref{cond1}) we can rewrite equation (\ref{cond2}) as
\begin{equation}
d_1  =  -x_c (c_1 + a_0 \ln x_c).
\end{equation}

Removing the step in the first derivative means that there will
be no delta function in the second derivative, merely a simple pole.
Throughout the rest of this
paper we will refer to these
solutions as zero-step solutions.
\citet{bal85} suggested that such solutions were modes in the corotation
region.  Using the criterion of a vanishing step in the first derivative
he was able to track the propagation of two such solutions
 into corotation as the amount of differential rotation
was increased. These solutions subsequently merged to give birth to a dynamically
unstable mode. Whilst it is not clear to us that these solutions are modes
in the traditional sense, they are clearly of interest if they
can give rise to dynamical instabilities.  They are also
notable because at the corotation point the Wronskian of the two solutions
being matched ($u_R$ and $u_L$) is zero. 

We have seen in this section that corotating solutions will in general
be singular at the corotation point.  At first sight this may seem
unphysical.  In solving the normal mode problem, however, it is implicitly
understood that our ultimate goal is the evolution of initial
data.  The physical time-varying solution is obtained by convolving the
solutions to the normal mode problem with the initial data and integrating
over the frequency range (an inverse Laplace transform). The
resulting integral solution will not be singular, and the continuous spectrum
will indeed have physical relevance \citep{van55}. In fact consideration of
the initial value problem leads us to suspect
that the zero-step solutions might exhibit behaviour that is physically
distinct from the rest of the continuous spectrum.  We are currently
investigating this issue (Watts et al, in preparation).

\section{Solutions on the boundary of the corotation region}

As discussed in Section \ref{uni}, the r-modes are the only 
non-trivial solutions in the case of uniform rotation.
An interesting question is whether these modes
enter the corotation region as the degree of differential rotation is
increased. In order to do this they must somehow 
develop corotation points. For the simple rotation laws
we consider in this paper there are two possibilities. 
Either the corotation point appears first at the poles, or at the equator.
In order for the mode to enter the corotation region, there
must exist a solution with a corotation point at the corresponding
endpoint. We will consider each possibility in turn.

\subsection{Corotation point at the pole}  
\label{pole}

First consider (monotonic) rotation laws for which the angular velocity
is lowest at the pole.  If modes are to enter corotation there must
be mode solutions with corotation points at the
pole. We can try to rule out such solutions by considering the 
corresponding behaviour at the pole.  This problem is 
easiest if we work with $U$, i.e consider equation (\ref{ymode}).

If the corotation point is at the pole we can search for a power series
expansion of the form 
\begin{equation}
U  =  \sum_n a_n (1-x^2)^{n+\gamma}.
\end{equation}
If $\tilde{\Omega}'(1)\ne 0$ then the indicial equation is
\begin{equation}
\gamma^2 = \frac{m^2}{4} + \frac{\tilde{\Omega}'(1)}{2 \Omega'(1)}.
\end{equation}

We now need to determine what values of $\gamma$ are permitted by the
boundary conditions.  It is clear from consideration of the physical
variables that we require $\delta v_\theta$ and $\delta v_\varphi$ to be zero
at the pole.  From the continuity equation we can see that the second
condition corresponds to requiring $dU/d\theta = 0$.  Now
$U  =  \delta v_\theta \sin\theta$ and
\begin{equation}
\frac{dU}{d\theta}  =  -(1-x^2)^{\frac{1}{2}}\frac{dU}{dx}. 
\end{equation}
We therefore have two conditions to meet at the pole: $U=0$ and $dU/dx$
less singular than $(1-x^2)^{-1/2}$.  These conditions immediately
rule out the possibility of $\gamma$ being imaginary.  For real $\gamma$ the 
conditions exclude the possibility of $\gamma$ being negative, moreover we
must have $\gamma > 1/2$.  Whether or not this
condition can be met will depend on the rotation law and the degree of
differential rotation.  

\subsection{Corotation point at the equator}
\label{equator}

Consider now rotation laws for which the rotation rate is lowest at the
equator.  If modes are to cross the corotation boundary there must
be mode solutions with corotation points at the equator.  Consider
equation (\ref{umode}).  If the corotation point is at the equator the
symmetry of the rotation law means that $m\Omega - \sigma$ will contain a
factor $x^2$.  As before we assume that $\tilde{\Omega}'(0)\ne 0$ and look
for power series expansions of the form 
\begin{equation}
u  =  \sum_n a_n x^{n+\gamma}.
\end{equation}
The indicial equation yields
\begin{equation}
\gamma = \frac{1}{2} \left[ 1 \pm \left(1 - \frac{8 \tilde{\Omega}'(0)}{\Omega''(0)
}\right)^{\frac{1}{2}}\right].
\end{equation}

If $\gamma$ is real we have two possible power series, one for each value
of $\gamma$ (which we call $\gamma_+$ and $\gamma_-$).  $\gamma_+ \ge  1/2$
and $\gamma_- \le 1/2$.  The $\gamma_- $ series has a
singular derivative, which would correspond to a singular $\delta v_\varphi$.
If $\gamma$ were imaginary we would have $\gamma = 1/2 \pm
i\tilde{\gamma}/2$, so once again the derivative would be singular.

Suppose that we permit singular solutions but restrict the degree of
singularity by requiring that the physical variables $\delta v_\theta$ and
$\delta v_\varphi$ be square integrable.  This is a reasonable restriction
as it enables us to define an energy measure.  For this restriction to
hold $u'$, and hence
$\delta v_\varphi$, must be less singular than $x^{-1/2}$. The only
permissible power series solution is therefore that with index $\gamma_+$.

If the maximum value of $\gamma_+$ is less than 1 then $u'(0)$ is singular for even
this solution.  This would
preclude the existence of even mode solutions with equatorial
corotation points (for even modes we expect $u'(0) = 0$).  In other words,
it appears even r-modes cannot cross into corotation for certain rotation laws.  The development
of equatorial corotation points in odd modes for these laws remains however
a possibility.

\section{Application to specific rotation laws}

The majority of the differential rotation laws examined
in the literature were introduced because of their mathematical
simplicity rather than being motivated by the anticipated physics.   
This is obviously fine as long as one is mainly interested 
in the properties of typical solutions, but it is 
clear that one ought to be somewhat cautious before drawing 
astrophysical conclusions from the relatively small set of
rotation laws that have been considered to date. 
Ideally one would like
to use a rotation law for a young neutron star that is grounded in
physics, emerging for example from simulations of core collapse
\citep{zwe97,dm02a}.  At this stage, however, we are 
trying to deal with the technical challenges posed by the 
differential rotation problem. It therefore makes sense to 
consider simple rotation laws expressed in terms of a
few parameters. Since we also
wish to compare our results to previous work we have selected
four rotation
laws discussed in other papers. We have also considered a fifth law that
exhibits two phenomena not seen for the other laws.

The first two rotation laws,  the j-constant and v-constant laws, have been
used extensively in the literature (see 
for example \citet{hac86}).
In particular, they were used in an examination of
stable non-corotating modes of a differentially rotating shell \citep{rez01}.    
The j-constant and v-constant laws are  examples of
rotation laws where the angular velocity is greatest at the pole.

The third rotation law was discussed in a paper on solar
oscillations \citep{wol98}.  This is one of the few laws encountered in  the literature
that is grounded in observations.
Wolff used this law to examine oscillations of a differentially
rotating shell.  This model caught our attention because Wolff's results
suggested a large number of corotating solutions that, as differential
rotation increased, combined with the more familiar r-modes
and disappeared, potentially giving rise to a dynamical instability.
However, Wolff  did not discuss either corotation points or
dynamical instabilities.  This law is also of
interest as it is an example of a rotation law where the angular
velocity is greatest at the equator.

Our fourth rotation law is a simplified version of the
Wolff law and is a particular instance of a law discussed by both 
\citet{bal85} and \citet{luy90a}. This law is interesting because 
(as we will see) it
exhibits both dynamical instabilities and regular solutions in corotation.

The fifth rotation law has angular velocity greatest at the pole, but
unlike the j-constant and v-constant laws, exhibits dynamical
instabilities and modes with equatorial corotation points.

All five laws obey the Solberg criterion for
dynamical stability to axisymmetric perturbations \citep{tas78}.  This criterion, which generalises the
Rayleigh condition for convective stability in an inviscid, incompressible
fluid, states that we must have
\begin{equation}
\frac{d}{d\varpi} (\Omega^2 \varpi^4) > 0,
\end{equation}
where $\varpi$, the cylindrical radius, is given by $\varpi =
\sin\theta = (1-x^2)^{1/2}$ (we have set the radius of the shell equal
to one).

During violent events such as core collapse or neutron star neutron star
merger it is conceivable that a star might achieve an intermediate rotation
state that does not obey the Solberg criterion.  The analysis conducted in
the next section for the shell will only be valid for a Solberg unstable
law if the timescale for growth of the
Solberg instability is low compared to the timescales associated with the
non-axisymmetric oscillations.  

\subsection{Method}
\label{method}
In our analysis of each rotation law we took the following steps
(in our view these would also be the natural steps in the analysis  
of other cases):

\begin{enumerate}
\item{Check whether the law obeys the Solberg criterion}
\item{Check whether the law has an inflexion point in the vorticity
($\tilde{\Omega}' = 0$).
This will determine whether the law can, in principle, have dynamical instabilities
or regular modes in corotation.}
\item{Solve for real-frequency modes outside corotation. This is the easiest 
calculation to do, and it will indicate whether modes are approaching the corotation 
region.}
\item{Solve for modes at the edge(s) of the continuous spectrum. This will 
provide information concerning whether modes cross into the 
corotation region. Furthermore, such solutions may (as we discuss later) 
indicate the onset of dynamical instability.}
\item{Search for regular corotating modes}
\item{Search for corotating modes with zero step in the first derivative at the
corotation point. These are solutions of the kind considered by \citet{bal85}.}
\item{Search for dynamically unstable modes developing at points where}
\begin{itemize}
\item{Stable real-frequency modes breach the instability criterion}
\item{Stable real-frequency modes merge}
\item{Zero-step corotating solutions merge}
\end{itemize}
\end{enumerate}

The notion that dynamical instabilities develop when modes merge is
supported by empirical evidence from other instability calculations.  The
dynamical bar mode instability that limits the angular velocity of rotating
stars sets in at points where two non-radial modes merge.  The radial
instability that sets the maximum mass of a neutron star also sets in
through the merger of two modes.   Theoretical support for this idea is given by
\citet{sch80}, who notes that whenever instability sets in along a sequence the
Hermitian eigenfrequency equation will have a double root. In addition, we
have already discussed how \citet{bal85} found
dynamical instabilities developing at the point where two zero-step
corotating solutions merged.

The numerical integrations necessary to find the modes were carried out
using a variable order variable step
Adams method.  A solution at one value of differential rotation was then
used as an initial estimate for the solution at a slightly increased or
decreased value.  In this way we were able to track modes through the
desired range of differential rotation.  

Modes without corotation points (real and complex frequencies) were located
by picking a trial frequency, integrating from pole to equator and then
iterating using the Newton-Raphson method until the equatorial boundary condition
appropriate to the mode symmetry was met.  A similar routine was used to
search for solutions on the corotation boundary, although in this case
because the location of the corotation point was fixed there was a fixed
relationship between frequency and amount of differential rotation.  

Corotating solutions were found by splitting the domain of integration into
two domains: $x_c < x \le 1 $ and $0 \le  x < x_c$.  As the first domain has
singularities at both ends we 
integrated away from both singularities and evaluated the Wronskian at a
matching point in the middle of the domain.  On the second domain we
integrated away from the singularity at $x=x_c$ and evaluated the function
and its derivative at the equator. The desired solutions were those for
which both the Wronskian on the first domain and either the
function or its derivative at the equator (depending on the symmetry of the solution)
vanished simultaneously.  For
both regular and general solutions in corotation the power series
expansions of Section \ref{discorot} were used to approximate the solution
at $x = x_c \pm \delta$ (where we used $\delta \sim 10^{-5}$ in the numerical 
calculations).  
When searching for regular solutions in corotation for which
$\tilde{\Omega}'(x_c)\ne 0$, we fix the arbitrary constant $a_0$ in the
power series expansion and vary the frequency.  For the 
general singular solutions the situation is a little more complicated.  In the power
series expansion for $x = x_c + \delta$ there are two free parameters,
$a_0$ and $c_1$.  By fixing these and imposing the matching conditions at
$x=x_c$ we completely determine the two free parameters in the power series
solution at $x = x_c - \delta$. In actuality the two desired functions 
become zero only  for particular values of both the frequency and the ratio
$c_1/a_0$ (we use $c_1/a_0$ rather than $a_0/c_1$ as our variable because there are
solutions where $c_1 = 0$).  We therefore fix $a_0$ and vary both the frequency and this ratio, using
a modified version of the Powell hybrid method to locate zeroes of the two functions.
Fortunately the frequencies and ratios of solutions turn out to be smooth functions of
the parameters that determine the amount of differential rotation.  

\subsection{Case 1: the j-constant rotation law}

The j-constant law for a shell takes the form
\begin{equation}
\Omega = \frac{\Omega_c A^2}{1 + A^2 -x^2}.
\end{equation}
The parameter $A$ controls the amount of differential rotation\footnote{In
the limit $A\rightarrow 0$ the j-constant law tends to
zero rotation.  We do not discuss this limit further in this paper and
hence assume $A>0$.},
uniform rotation ($\Omega = \Omega_c$) being achieved in the limit as $A\rightarrow
\infty$.  The rotation rate for this law is lowest at the
equator.
We get
\begin{equation}
\tilde{\Omega}' = \frac{2\Omega_c A^4 (1+A^2 + 3x^2)}{(1+A^2 - x^2)^3}.
\end{equation}
Since
$\tilde{\Omega}'$ is always positive for real $A$
there will be no
dynamically unstable modes or  regular corotating modes for this
rotation law.

\begin{figure*}
\includegraphics[height = 5.5cm,clip]{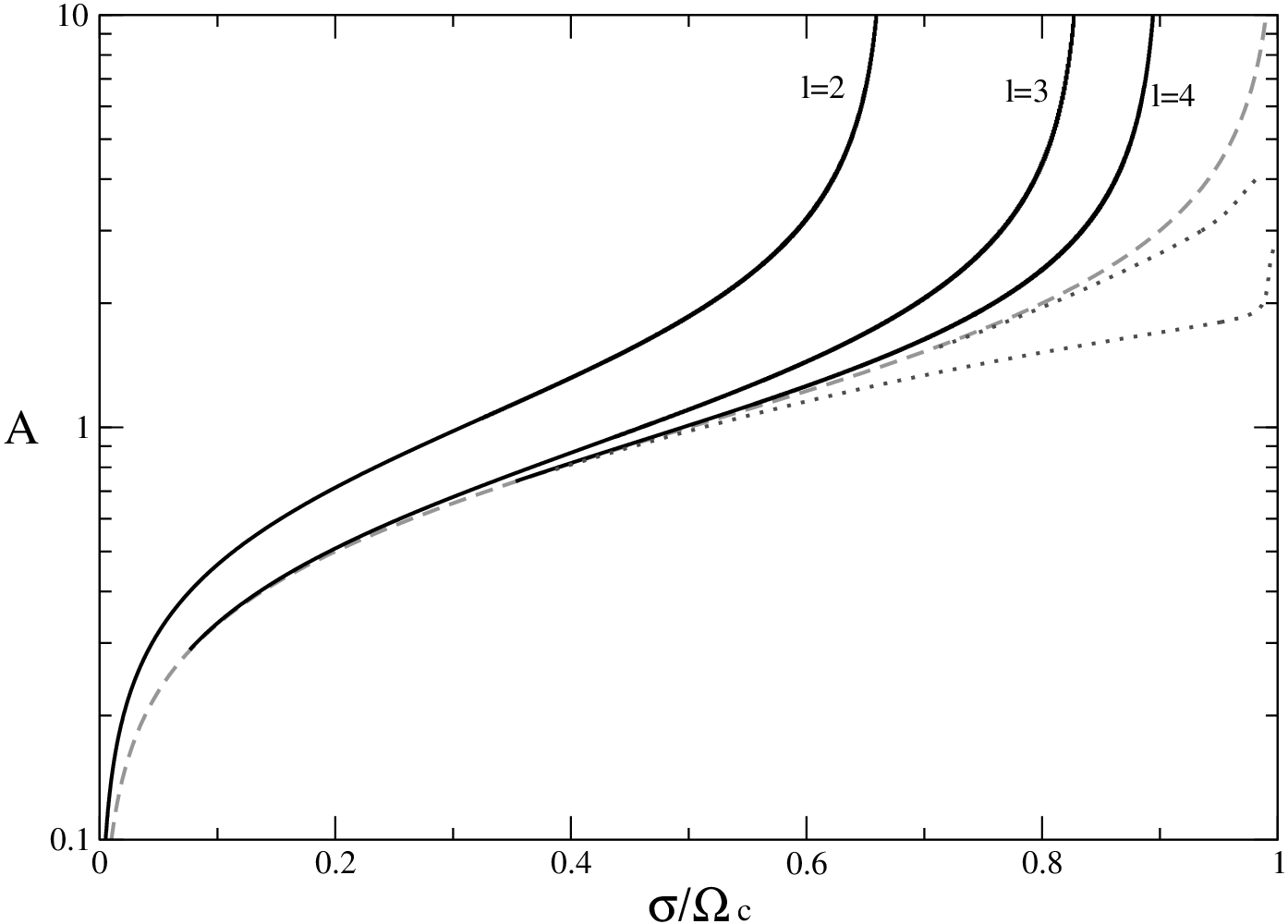}
\hspace{1cm}
\includegraphics[height = 5.5cm,clip]{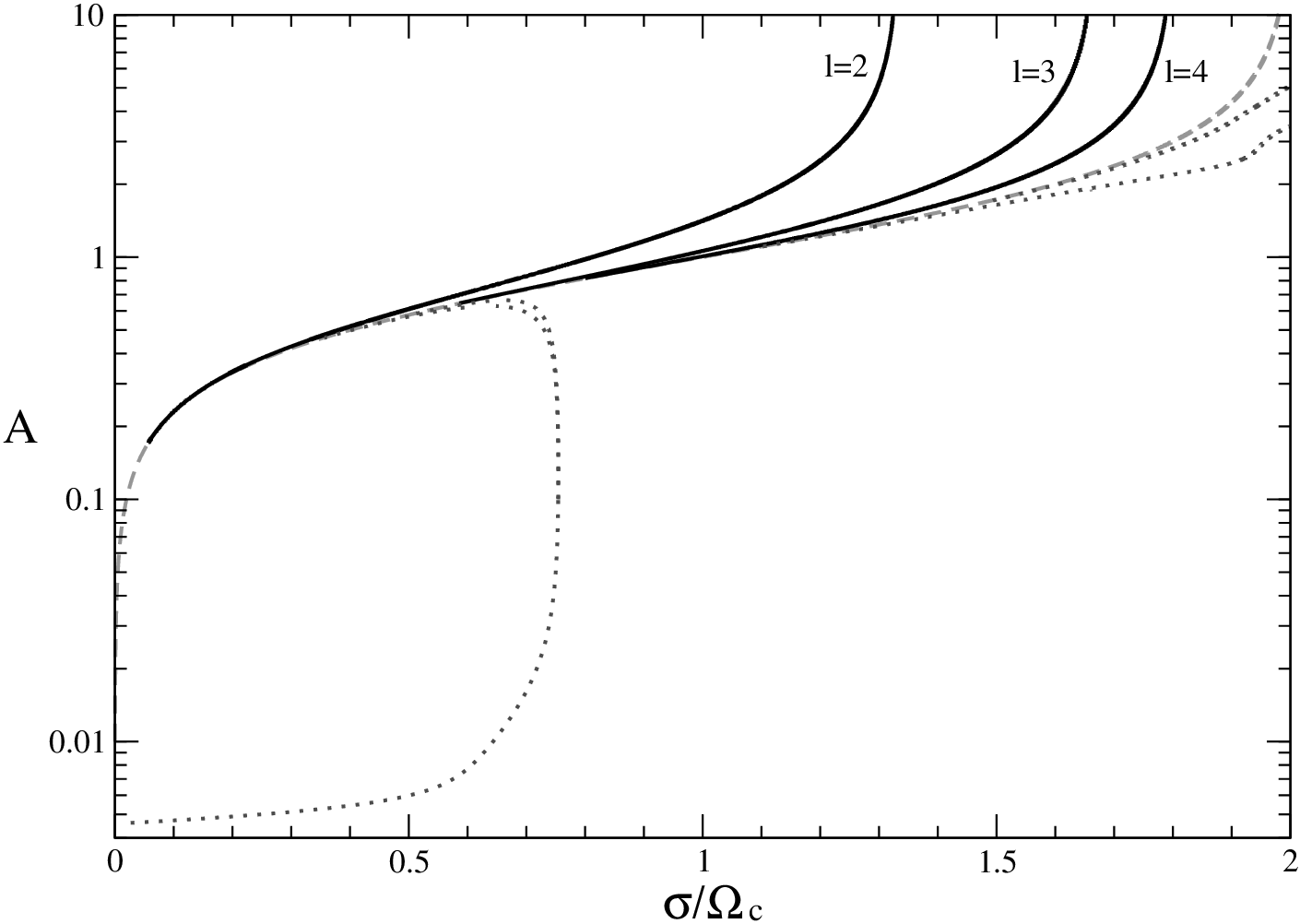}
\caption{Results for rotation law 1:  
$A$ determines the degree of differential rotation, rotation being
uniform in the limit $A\rightarrow \infty$. The plot on the left shows $m=1$, that on the right $m=2$.  Solid
lines:  real-frequency modes outside corotation.  Dashed
line:  The lower boundary of the corotation region -  solutions to the right of
this line are in corotation.  Dotted lines:  real-frequency solutions with
zero step in the first derivative at the corotation point. The values
of $l$ given refer to the uniform rotation limit $A\rightarrow \infty$, in
which, as discussed in section \ref{uni}, there is one r-mode
solution for each value of $l$.}
\label{j}
\end{figure*}

Figure~\ref{j} shows the frequencies of a selection
of modes and
corotating zero-step solutions for $m=1$ and $m=2$, respectively.  Note
that for any value of $A$ there is a singular solution at every frequency
within the corotation band.  In the Figures for this and the other rotation
laws, however, we only show the
zero-step solutions.  

The frequencies derived for modes outside corotation confirm the
results of \citet{rez01}.  As the amount of differential rotation
increases, the frequencies of the r-modes decrease and approach 
corotation. The code crashes as
$\sigma \rightarrow m\Omega$ at the equator, crashing at
larger values of $A$ for modes that at uniform rotation have larger
 $l$.  Modes that have very large $l$ in uniform
rotation are not shown, but follow the general trend illustrated in
the figures.  Whether modes continue to exist very close to the corotation
boundary as the degree of differential rotation is increased still further
is difficult to say.  We have however tracked modes very close to the
corotation boundary and our results suggest that modes outside corotation
cease to exist below a certain value of $A$.  Disappearance of modes as a
system parameter changes is not unexpected.  Several examples of
eigenvalues vanishing when parameters change are given in
\citet{kat95}.  

The question of whether modes can enter the corotation region by developing
equatorial corotation points was discussed in section \ref{equator}.
For this law the maximum value of $\gamma_+$ is 1.  This means that,
analytically, we were able to rule out any even solutions with equatorial
corotation points
for any value of $A$, and any odd solutions for  $A > 0.354$.  This was
confirmed by our numerical results.  For $A<
0.354$ we were not able to rule out analytically the possibility of odd
r-modes crossing the boundary.  Our numerical results suggest however that
there are no odd r-mode solutions existing below $A\approx 0.36$.  For this
rotation law, therefore, we find no evidence that r-modes cross into
corotation.   

\subsection{Case 2:  The v-constant rotation law}

Another law that has been examined in the literature is the v-constant
rotation law.  There are however two different laws with this name
 in the literature.  The first, used by \citet{hac86} and \citet{kar02}, takes the form

\begin{equation}
\Omega = \frac{\Omega_c A}{(A^2 + 1 - x^2)^{1/2}}.
\end{equation}

The second, used by \citet{eri85}, \citet{kar01}, and 
\citet{rez01}, takes the form

\begin{equation}
\Omega = \frac{\Omega_c A}{A + (1-x^2)^{1/2}}.
\end{equation}

In both cases uniform rotation ($\Omega = \Omega_c$) is reached in the
limit $A \rightarrow \infty$.  For the first law it is not possible to have
$\tilde{\Omega}' = 0$, so we can rule
out the possibility of dynamically unstable modes or stable modes 
within corotation.  For the second,  $\tilde{\Omega}'$ can be zero
for $A>1$ or $A<1/2$, so we cannot a priori rule out the possibility of
unstable modes.  No instabilities were found, however, by our numerical
analysis.  In fact the modes and continuous spectrum for the two v-constant
laws behave qualitatively identically to the modes and continuous spectrum
of the j-constant law.  For this reason we have not included figures of the
results.

\subsection{Case 3:  Wolff's solar rotation law}

Wolff's solar rotation law \citep{wol98} takes the form 
\begin{equation}
\Omega = 2\pi(454.8 -60.4\beta x^2 - 71.4\beta x^4).
\end{equation}
The parameter $\beta$ measures the amount of differential rotation,
uniform rotation corresponding to $\beta = 0$.  Wolff considers the range
$0 < \beta < 1$ but we will consider an upper limit of $\beta = 2$ as
there are interesting phenomena in this range.  The rotation rate for
this law is lowest at the pole.  
We get
\begin{equation}
\label{wolffomt}
\tilde{\Omega}' =  2\pi(909.6 + 120.8\beta + 132\beta x^2 - 2142\beta x^4).
\end{equation}
In this case $\tilde{\Omega}'$ has a zero on the domain $0 \le x \le 1$ for
$\beta > 0.4815$.  Both dynamical instabilities and regular
solutions in corotation are therefore possible.

\begin{figure*}
\includegraphics[height = 5.5cm,clip]{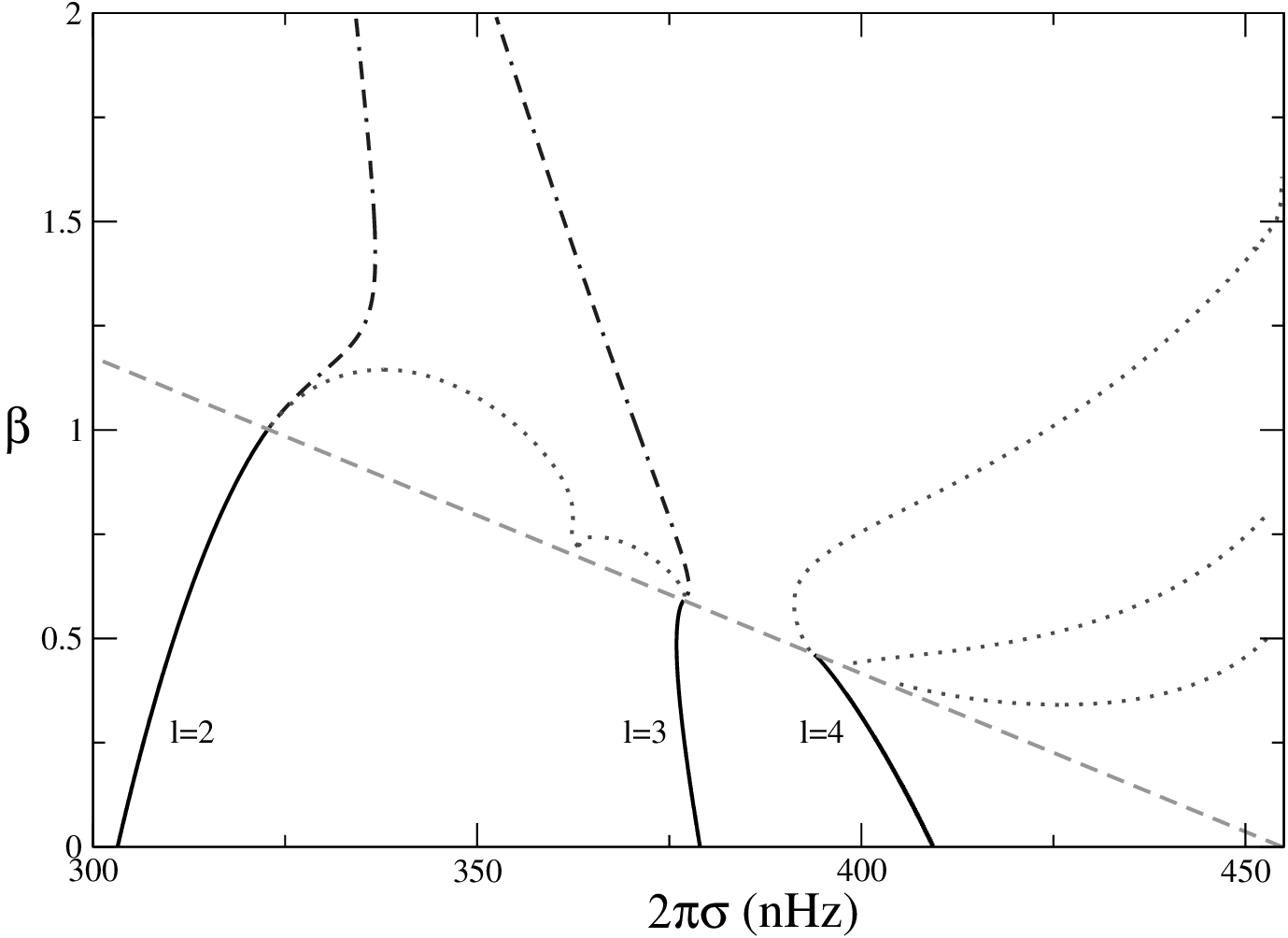}
\hspace{1cm}
\includegraphics[height = 5.5cm,clip]{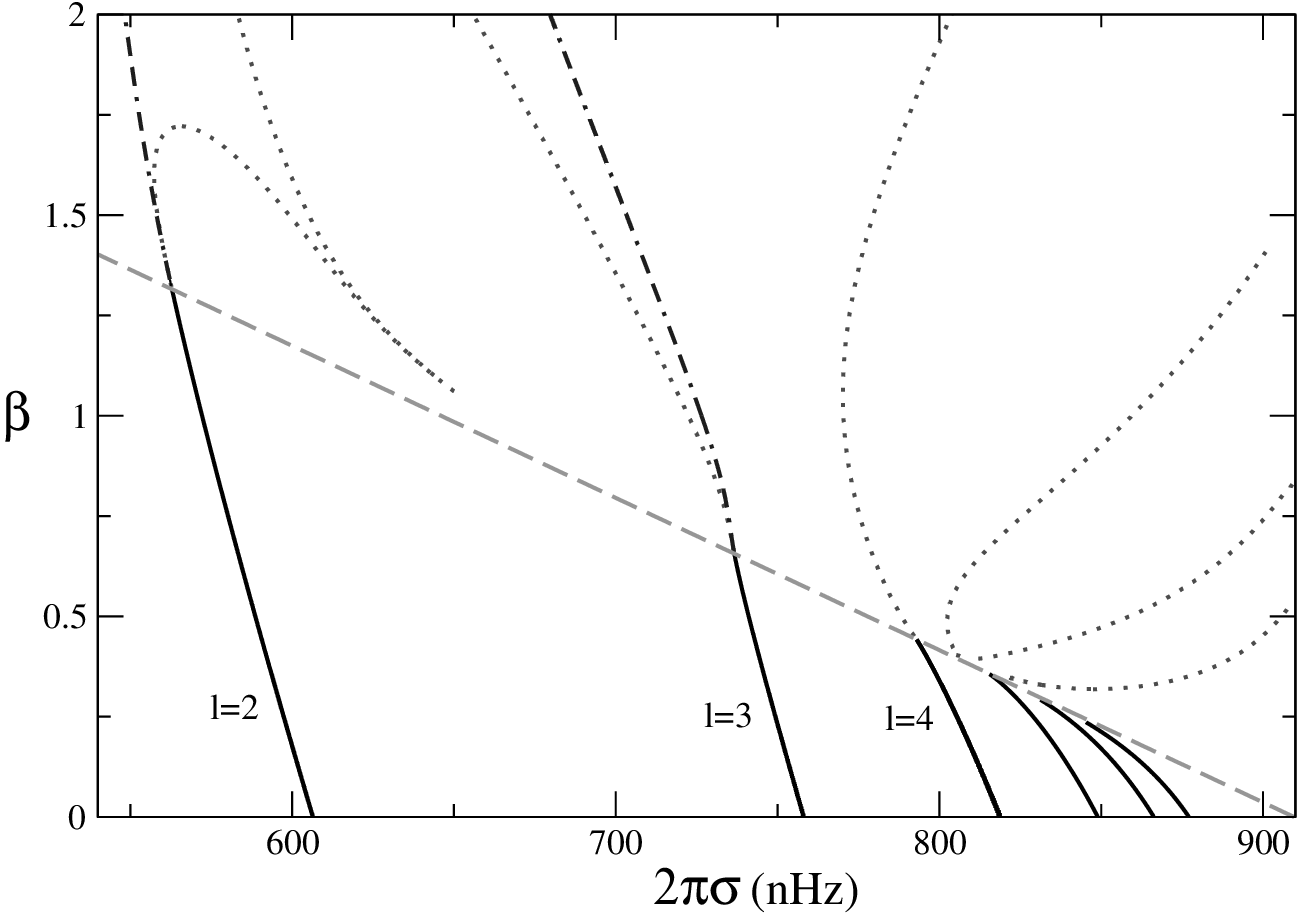}
\caption{Results for rotation law 3: $\beta$ determines the degree of differential rotation,
rotation being uniform when $\beta = 0$.  The plot on the left shows $m=1$, that on the right $m=2$.
Solid lines:  real-frequency modes outside corotation.  Dashed
line:  The lower boundary of the corotation region -  solutions to
the right of
this line are corotating.  Dash-dot lines:  Real part of frequency for
dynamically unstable modes. Dotted lines:  real-frequency solutions with
zero step in the first derivative at the corotation point. For dynamical
instabilities to exist, we require $\beta > 0.4815$. The values
of $l$ given refer to the uniform rotation limit $\beta = 0$, in
which, as discussed in section \ref{uni}, there is one r-mode
solution for each value of $l$.}
\label{w}
\end{figure*}

Figure~\ref{w} shows the frequencies of a selection
of modes and
zero-step corotating solutions for $m=1$ and $m=2$.  As the amount of
differential rotation increases, the frequency of the
each r-mode shifts and in due course all modes encounter the corotation
boundary.  Modes of low $l$ appear to approach the boundary directly,
modes of higher $l$ seem to approach tangentially.  Searching for
mode solutions on the corotation boundary, as described in section~\ref{pole}, 
we find valid solutions at the points where the $l=(2,3)
, m=1$ and $l=(2,3,4), m=2$ modes encounter the
boundary, suggesting that these modes can develop
corotation points\footnote{Figure \ref{w} appears to suggest that the $l=4, m=1$ mode crosses the
boundary.  There is however no physically valid solution on the boundary.
Resolving this region in more detail it is possible to see the $l=4$
mode and the corresponding zero-step corotating solution approaching the
boundary tangentially but not crossing.}. As $m$ increases so does the number of valid
solutions on the corotation boundary.  For $m=3$ and $m=4$ there are
valid solutions at the point where the four lowest $l$ modes encounter
the corotation boundary.  Modes of larger $l$, which appear to
approach tangentially, do not seem to cross the boundary.  

A numerical search for modes that are regular in corotation but for
which the corotation point does not coincide with the zero of $\tilde{\Omega'}$
proved fruitless.  For the range of $\beta$ examined there were no
valid solutions on the domain $x_c < x < 1$ (see the discussion in Section \ref{method}).  

What about the 
possibility of regular solutions, with corotation points at the point where
$\tilde{\Omega}' = 0$?  Such solutions could only exist at one frequency for
a given $\beta$, this frequency being a smooth function of $\beta$.  If we
represent the solution as a sum over spherical harmonics, $U = \sum C_l^m
Y_l^m$, it is possible to derive a recursion relation for the coefficients
$C_l^m$.  A numerical search for these modes yields nothing, however, and a
numerical examination of the coefficients $C_l^m$ shows why.  The coefficients are
divergent, with $C_l^m \rightarrow \infty$ as $l\rightarrow \infty$. There are
therefore no regular modes in corotation for this law.  

Continuous spectrum solutions with zero-step in the first derivative
are present for this rotation law.  Such solutions are generated at the point where
real-frequency modes appear to enter corotation (i.e. where there are
valid solutions on the corotation boundary).  There are also many such solutions present at
low differential rotation which, in contrast, seem to approach both
edges of the corotation region tangentially.

\begin{figure*}
\centering
\includegraphics[height = 11cm,clip]{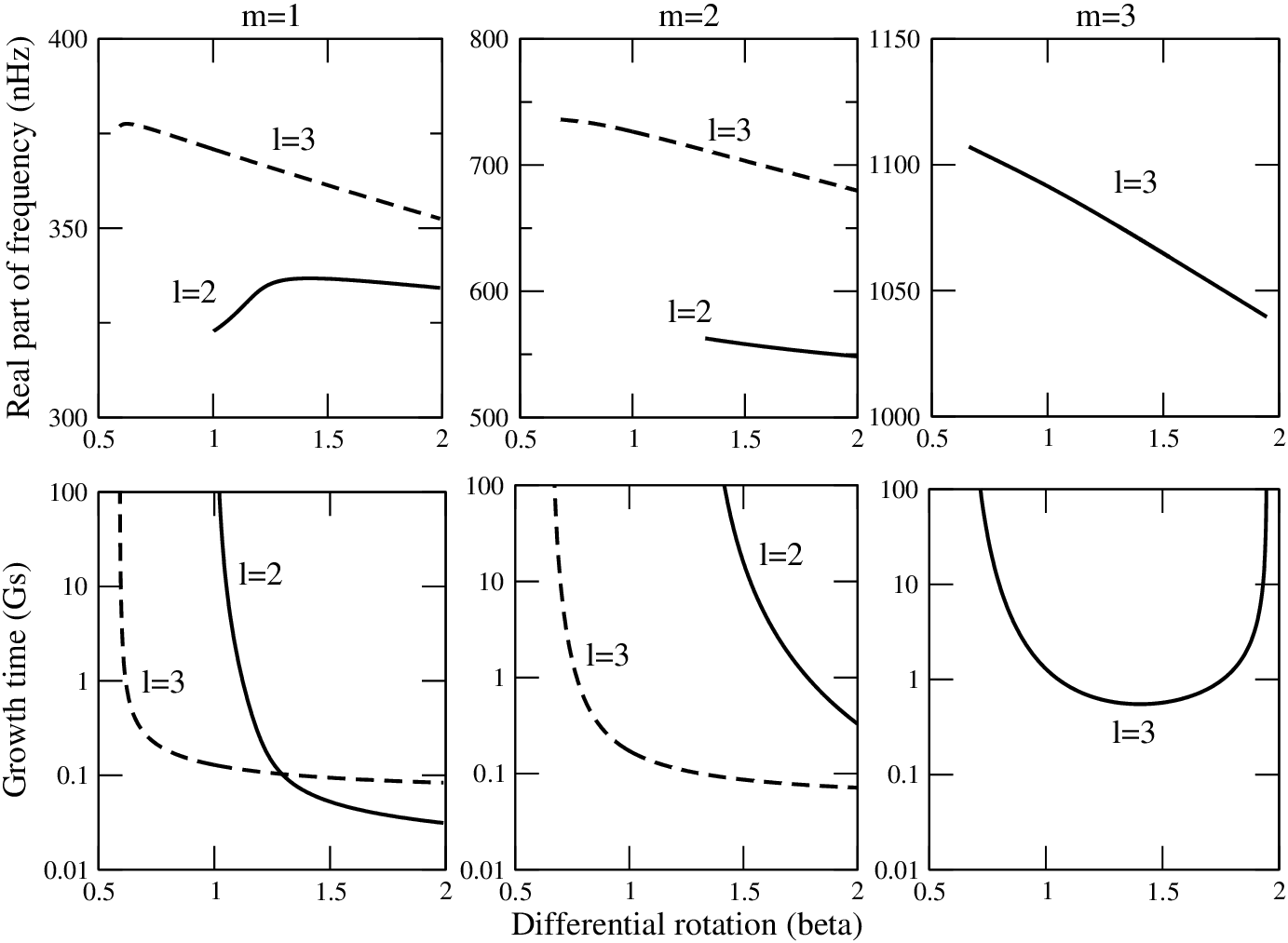}
\caption{Dynamically unstable modes for law 3.  The upper three frames
show the real parts of frequency, while the lower frames show the corresponding
growth times.}
\label{wd}
\end{figure*}

Dynamical instabilities are found to develop at the point where modes
encounter the corotation boundary at values of $\beta$ for which  $\tilde{\Omega}'$
has a zero on the shell (in accordance with the instability
criterion derived in Section \ref{dyncon}).  
Modes that encounter the corotation boundary below the
threshold value of $\beta$ do not exhibit instabilities.  There are
five unstable modes, corresponding to  $l=(2,3),m=1; l=(2,3) m=2$ and $l=3,
m=3$.  We find no instabilities for modes of higher $l$ or $m$.
Frequencies and growth times for these unstable modes are
shown in figure~\ref{wd}.  The behaviour of the $l=m=3$ mode is
particularly interesting as it indicates that the mode becomes stable again
for high differential rotation. The $l=3, m=2$ mode exhibits
similar behaviour at higher $\beta$.  

We can see from equations (\ref{ins2}) and (\ref{wolffomt}) why we
should not expect to find any unstable modes at high $\beta$.   As
$\beta$ increases, the negative part of $\tilde{\Omega}'$ will come to dominate the integrand in
equation (\ref{ins2}).  Beyond a certain threshold value it will no longer be
possible to satisfy the integral equation.  The precise value of the threshold will
depend on $U$ but all modes will reach this point for some $\beta$.  

The results reported in this paper agree with those reported by Wolff for
real-frequency modes outside corotation.  However, they differ dramatically
for solutions with corotation points.  In addition, we find dynamical
instabilities not mentioned by Wolff.  The discrepancy between the results
likely originates with Wolff's perturbation method and with his choice of
basis eigenfunctions.  He uses not only the standard r-mode eigenfunctions
discussed in section \ref{uni} but a set of toroidal geostrophic eigenfunctions
of single $l,m$.  But the geostrophic solutions form a
degenerate set in uniform rotation and cannot be classed as
eigenfunctions. As such they cannot be used as a basis for perturbation.

We can compare the results from our investigation with
measurements of the degree of differential rotation at various depths in
the Sun \citep{how00}.  The degree of differential rotation is highest in
the surface layers and the convective envelope, where values of $\Omega(0)/2\pi = 450$~nHz and
$\Omega(\cos(\pi/6))/2\pi = 370$~nHz are reported. This would correspond to having $\beta
\approx 0.9$.   At this level in our model we see dynamically unstable
modes, the fastest having a growth time of just a few years. 

Our model is very simplistic.  It does not, for example, include viscosity,
which would be expected to shift the points of marginal stability to a
higher degree of
differential rotation.  However, in the outer layers of the Sun the degree
of differential rotation is almost independent of the
radius, suggesting that a shell model may not be a bad approximation. In
this case a dynamical instability of the type found in this model could be
involved in limiting the degree of differential rotation observed in the
Sun. See also \cite{dzh02} for a discussion of various Solar properties that might be
explained by the action of unstable r-modes with growth times of
$\sim 10 - 1000$~
years.  Whether dynamical instabilities of the type found in this paper
could play a similar role is an open question.
  
\subsection{Case 4: A simple quadratic rotation law}

\begin{figure*}
\includegraphics[height = 5.5cm,clip]{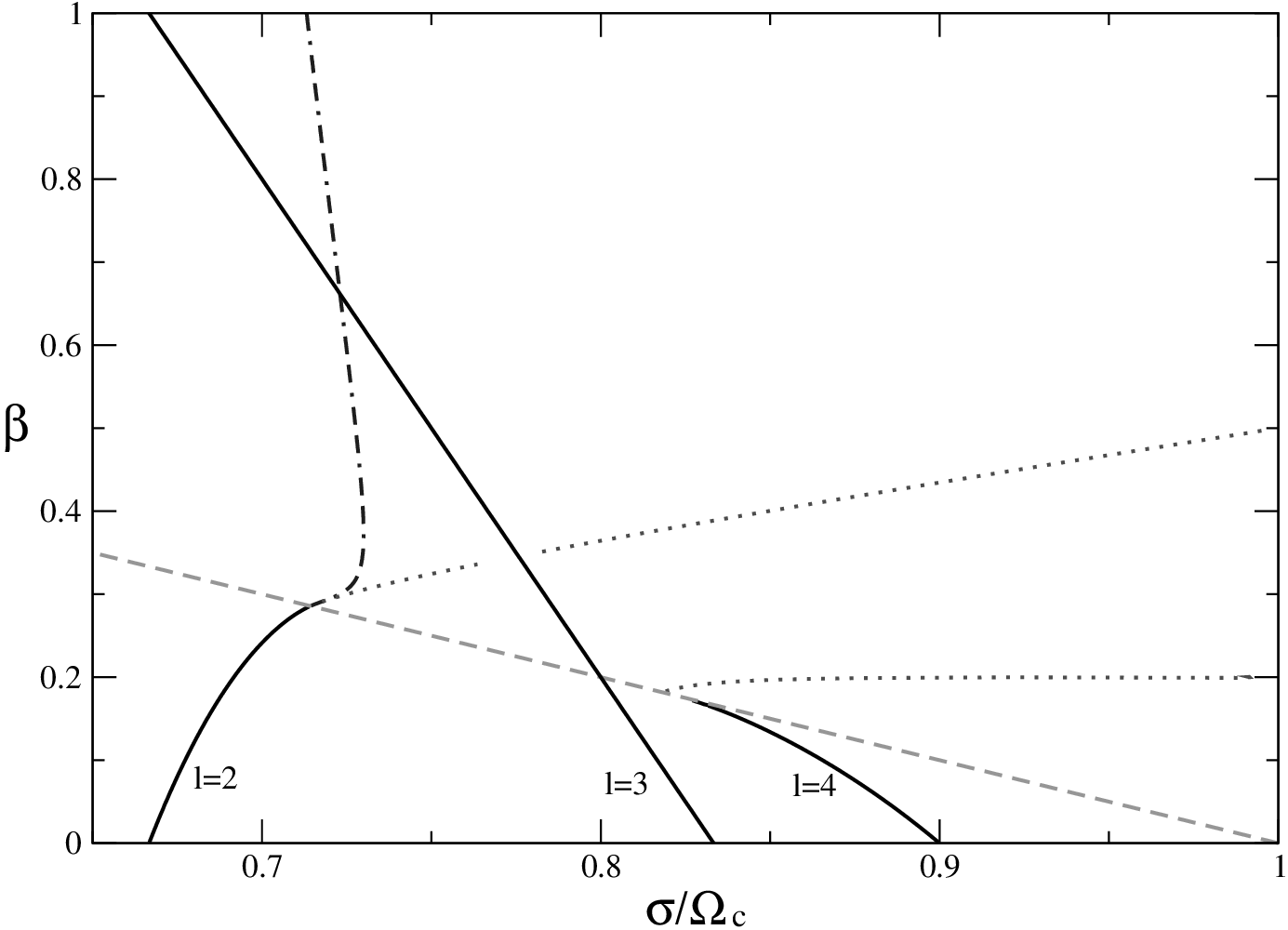}
\hspace{0.5cm}
\includegraphics[height = 5.5cm,clip]{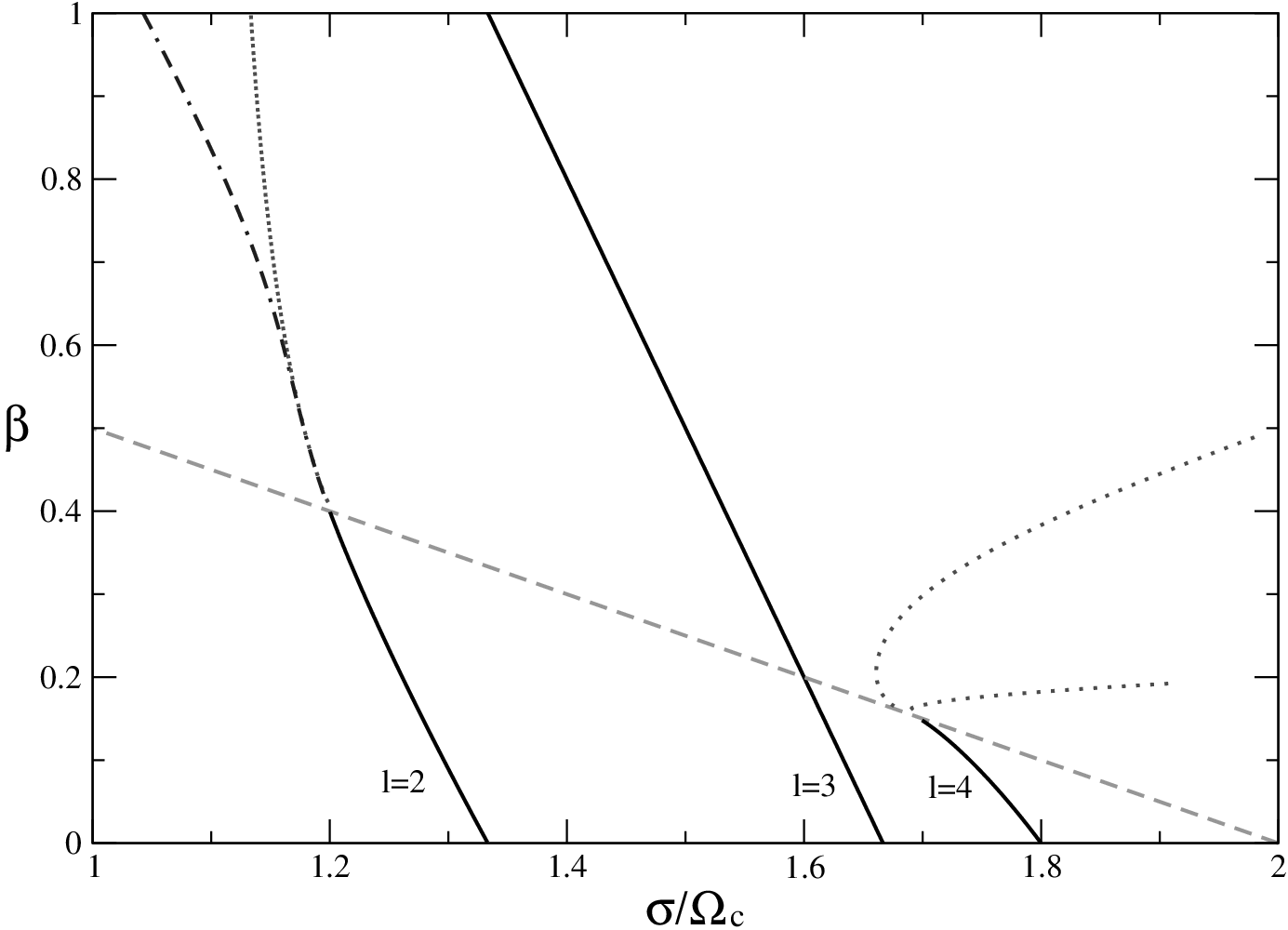}
\caption{Results for rotation law 4:  $\beta$ determines the degree of differential rotation,
rotation being uniform when $\beta = 0$. The plot on the left shows $m=1$, that on the right
$m=2$. Solid lines:  real-frequency modes outside corotation and
regular modes inside corotation.  Dashed
line:  The lower boundary of the corotation region -  solutions to
the right
this line are corotating.  Dash-dot lines:  Real part of frequency for
dynamically unstable modes.  Dotted lines:  real-frequency solutions with
zero step in the first derivative at the corotation point. For dynamical
instabilities to exist we require $\beta > 0.2$. The values
of $l$ given refer to the uniform rotation limit $\beta = 0$, in
which, as discussed in section \ref{uni}, there is one r-mode
solution for each value of $l$.}
\label{n}
\end{figure*}

The fourth rotation law we have considered takes the form
\begin{equation}
\Omega = \Omega_c(1-\beta x^2).
\end{equation}
Here the parameter $\beta$ measures the degree of differential rotation and we
have examined the range $0 \le \beta \le 1$.  Uniform rotation ($\Omega =
\Omega_c$) is recovered if
$\beta=0$.  As with Wolff's law the rotation rate for this law is
greatest at the equator.
We get
\begin{equation}
\tilde{\Omega}' = 2\Omega_c(1+\beta - 6\beta x^2).
\end{equation}
From this we see that
$\tilde{\Omega}'$ has a zero on the domain if $\beta \ge 0.2$. Both dynamical instabilities and regular
solutions in corotation are therefore possible. Figure \ref{n} shows the frequencies of a selection
of modes and
zero-step corotating solutions for $m=1$ and $m=2$.

Just as for Wolff's law, modes of low $l$ approach the corotation
boundary directly as differential rotation is increased.  Modes of
higher $l$ tend to approach tangentially.  And in accordance with
this there are valid solutions
on the corotation boundary for the $l=(2,3), m=1$ and $l=(2,3), m=2$
modes. 

The $l=3$ mode is a little different however, as it passes
straight into corotation as a regular mode.  To see why this happens consider the
following argument.  For this rotation law we can write
\begin{equation}
m\Omega - \sigma = -m\beta\Omega_c\left[x^2 - \frac{(m\Omega_c-\sigma)}{m\Omega_c\beta}\right],
\end{equation}
and
\begin{equation}
\tilde{\Omega}' = -12\beta\Omega_c\left[x^2 - \frac{1+\beta}{6\beta}\right].
\end{equation}
As discussed in section \ref{sreg}, the zeroes of these two functions 
may coincide, making the governing equation regular and giving rise to
regular solutions in corotation. This
will occur if
\begin{equation}
\sigma = \frac{m\Omega_c(5-\beta)}{6}.
\end{equation}
In this case the governing equation (\ref{ymode}) becomes
\begin{equation}
\label{srand}
\left[(1-x^2) U'' - 2xU' - \frac{m^2 U}{1-x^2}\right] + 6U = 0.
\end{equation}
This equation is simply Legendre's equation for $l=3$.  Thus there is a
pure $l=3$ mode that corotates but is completely regular.  Just as for
Wolff's law, however, there are no regular corotating solutions of the type given in equation
(\ref{ureg}).

\begin{figure*}
\centering
\includegraphics[height = 6cm,clip]{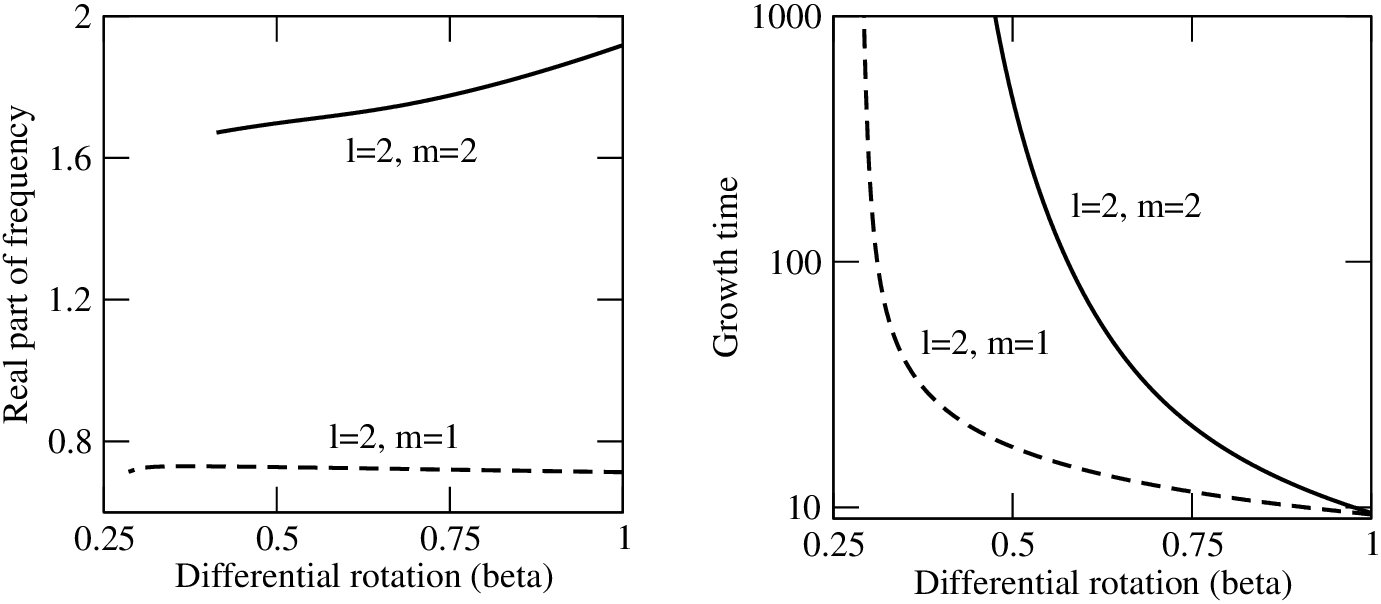}
\caption{Dynamically unstable modes for law 4.  The left frame
shows the real part of the frequency in dimensionless units
($\mathrm{Re}\sigma/\Omega_c$).   The right frame shows the growth
time in dimensionless units ($\Omega_c/\mathrm{Im}\sigma$).} 
\label{ad}
\end{figure*}

Only the $l=2,m=1$ and $l=2,m=2$ modes become dynamically unstable at
the point where they encounter the corotation boundary.  These are the
only modes that hit the boundary after passing the threshold value of
$\beta$ for which $\tilde{\Omega}'$ first has a zero on the shell.
Frequencies and growth times for both modes are shown in figure~\ref{ad}.  

\subsection{Case 5: A simple trigonometric rotation law}

The final rotation law we have considered is 

\begin{equation}
\Omega = \Omega_c \left(1 - \beta \cos x\right)
\end{equation}

\begin{figure*}
\centering
\includegraphics[height = 6cm,clip]{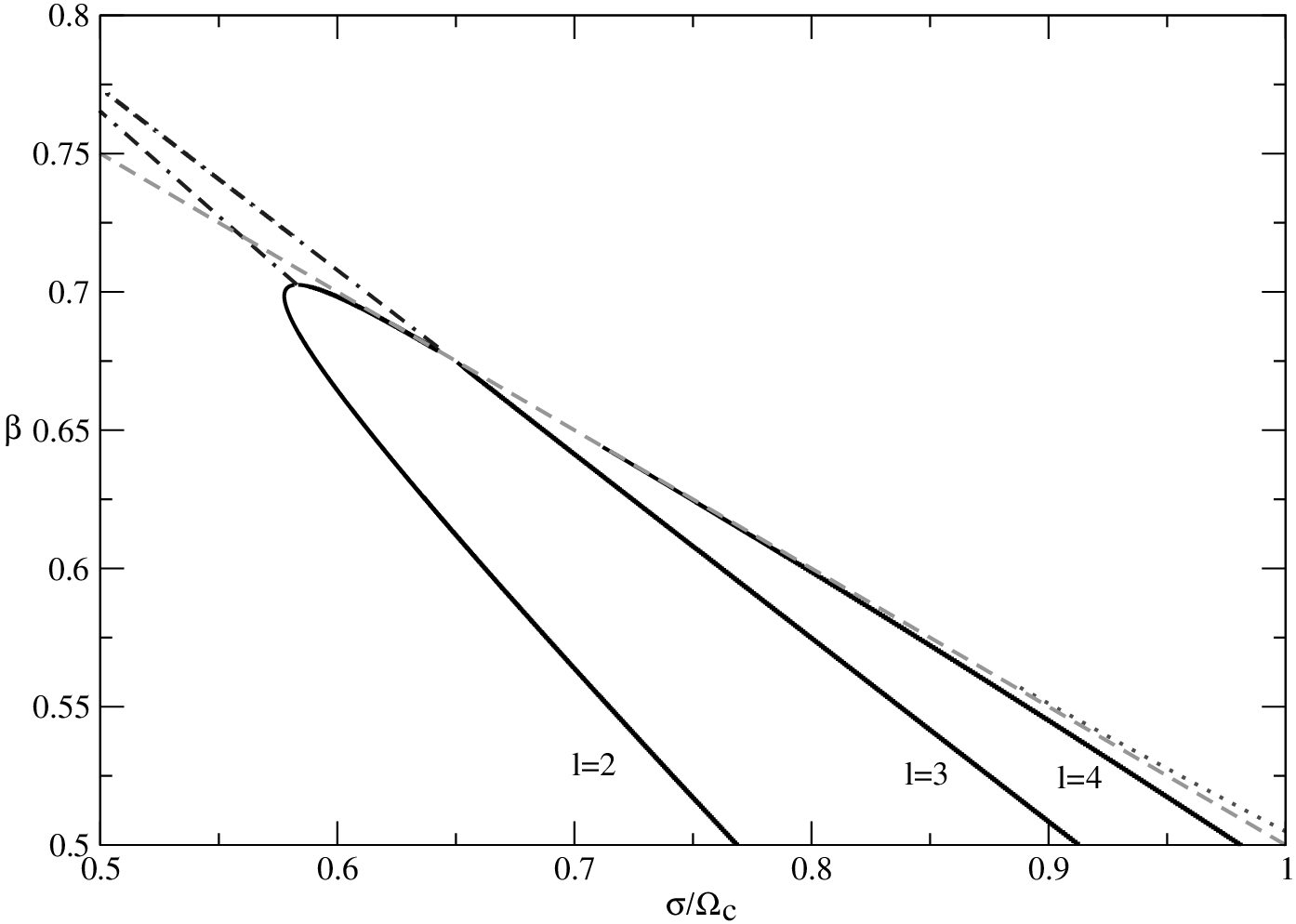}
\caption{Results for rotation law 5 for $m=2$: $\beta$ determines the degree of
differential rotation, rotation being uniform when $\beta = 0$. Solid
lines: real-frequency modes outside corotation.  Dashed line: The lower
boundary of the corotation region - solutions to the right of this line are
corotating.  Dash-dot lines:  Real part of frequency for dynamically
unstable modes.  Dotted lines:  Real frequency solutions with zero step in
the first derivative at the corotation point.  For dynamical instabilities
to exist we require $\beta > 2/3$.  The values of $l$ given refer to the
uniform rotation limit $\beta = 0$, in which, as discussed in section
\ref{uni}, there is one r-mode 
solution for each value of $l$.} 
\label{anna}
\end{figure*}

The parameter $\beta$ measures the degree of differential rotation and we
have examined the range $0\le \beta \le 1$.  Uniform rotation ($\Omega =
\Omega_c$) is recovered if $\beta = 0$.  The rotation rate for this law is
greatest at the pole, as it is for the j-constant and v-constant laws.
What is different about this law, however, is that $\tilde{\Omega}' = 0$
for $\beta > 2/3$, which could permit dynamical instabilities. 
Figure \ref{anna} shows mode frequencies for $m=2$ for a reduced range of
the differential rotation parameter $\beta$.  The figure shows
interesting phenomena that we do not see for any of the other rotation
laws. Modes crossing the corotation boundary for this law would have to develop
equatorial corotation points, as discussed in section \ref{equator}.  For
this law, the maximum value of $\gamma_+$ is 2.  Modes of odd symmetry can
in principle
cross the boundary if $\beta > 16/25$, modes of even symmetry if $\beta >
2/3$.

Let us start by considering the $l = 3$
mode.  As the amount of differential rotation is increased, the frequency
approaches the corotation boundary.  The mode crosses the corotation boundary above the
instability threshold value of $\beta$, and above this point
continues to exist as a dynamically unstable mode.  An odd zero-step
solution also develops at this point, but this is not shown in Figure
\ref{anna}, as the frequency is very close to the real part of the frequency of
the dynamically unstable mode. 

The situation for the even modes is slightly different.  Although the modes
could in principle cross the corotation boundary, we find that the
$l=2,m=2$ and $l=4,m=2$ modes
merge outside corotation, becoming dynamically unstable.  Such a route to
instability is not seen for the other rotation laws examined.   With
reference to Figure \ref{anna}, note that we
were unable to track the $l=4$ mode for $0.644< \beta < 0.678$ because the
frequency was too close to the corotation boundary.  We are however confident
that it is this mode that is merging with the $l=2$ mode, for three reasons.
Firstly, the merging mode has even symmetry (as we would expect).
Secondly, all even modes with $l>4$
approach the corotation boundary at lower values of $\beta$.  The $l=4$
mode is the only other even mode apart from $l=2$ that appears to persist
to high $\beta$.  Thirdly, the merging mode is not emerging from corotation,
as the only valid solution on the corotation boundary for $m=2$ occurs at
the point where the $l=3$ mode crosses.  The most reasonable
conclusion is that the $l=4$ mode continues to exist just outside
corotation for $0.644< \beta < 0.678$ before moving away from the boundary
to merge with the $l=2$ mode at higher $\beta$.  

\section{Summary}

The oscillations of neutron stars are thought to be a promising source of
gravitational waves.  In order to develop accurate models of gravitational
wave emission we need to develop a full understanding of the dynamics of
these stars.  One factor that has been neglected in many previous studies
is differential rotation.  Neutron stars are thought to be born with
differential rotation, and there are several plausible mechanisms that could
sustain differential rotation beyond birth.  Differential rotation
introduces the possibility of dynamical shear instabilities \citep{pap84,pap85}.  In addition,
the perturbation equations
of differentially rotating inviscid systems are complicated by the presence of
singularities at points where the pattern speed of a mode equals the local
angular velocity of the background flow.  These singularities mean that differentially rotating inviscid
systems possess a continuous spectrum in addition to discrete modes.  To
begin to understand this problem we have in this paper analysed the normal
mode problem for a toy model, a differentially rotating shell.
Although the model is simplistic, it exhibits many of the key features of
the more complex three-dimensional problem.  

In the uniform rotation limit the shell problem admits the standard r-modes as solutions.  As
the degree of differential rotation increases the frequencies of these
modes gradually approach the lower edge of the expanding corotation band ($m\Omega_{min} < \sigma <
m\Omega_{max}$).  Modes that cross the corotation boundary are found to be
rare.  Indeed, we can rule out completely the crossing of even modes for certain
rotation laws.  For other rotation
laws and symmetries we find crossings only for modes that have low $l$ in
the uniform
rotation limit. All other modes approach the boundary tangentially and
appear to cease to exist when the degree of differential rotation exceeds a 
threshold value.  Modes that have higher $l$ in the uniform rotation limit
cease to exist at lower degrees of differential rotation.

For frequencies in the corotation band the shell problem admits a continuous spectrum of
corotating solutions with discontinuous first derivatives.  Although such
solutions may appear at first sight unphysical, they will have physical
significance when we consider the initial value problem.  In general,
corotating solutions have both a logarithmic singularity and a finite step in
the first derivative at the corotation point.  For certain special
frequencies, however, the step in the first derivative vanishes.  In
differentially rotating cylinders these zero-step
solutions can merge to give rise to
dynamical instabilities \citep{bal85},
but we observe no such behaviour in our problem for the rotation laws
examined.  We cannot rule out the possibility, however, for other rotation
laws.  

The character of the modes that cross into corotation varies. At all points where modes
cross the corotation boundary we observe the development of zero-step
corotating solutions (except in one unique case, discussed below).  At some
points we also observe the development of dynamically unstable modes.  A
necessary (but not sufficient) condition for dynamical instability is that
$\tilde{\Omega}'$ be zero somewhere on the shell.  For the rotation laws
examined this condition is met only when the degree of differential
rotation exceeds a threshold value.  All modes crossing the corotation
boundary above this threshold become dynamically unstable.  We also
observe, for one rotation law, the development of a dynamical
instability via the merger of two stable modes outside corotation.  Only in one
exceptional case for the simplest rotation law did we observe a mode crossing the corotation boundary and
continuing to exist as a regular stable mode.  Such regular corotating
modes are unlikely to exist for more realistic rotation laws.  

Before moving on to the three-dimensional problem, we intend to investigate
two related problems for the simple shell model. The first factor of
interest is the effect of viscosity, an essential component of any
realistic model. If we
consider the Navier-Stokes equations instead of the Euler equations the
vorticity equation for the stream function becomes
\begin{eqnarray}
\label{navier}
(\sigma-m\Omega) \nabla^2_\theta U + {m \partial_\theta
\tilde{\Omega} \over \sin \theta} U  
- \frac{i\nu}{\rho R^2} \Big[ \nabla^2_\theta(\nabla^2_\theta  U) & + & {}
\nonumber \\ 
\frac{m^2}{\sin^4\theta}\left(\sin \theta \cos\theta \partial_\theta U -
(\cos^2\theta + m^2)U\right)
\Big] &  = & 0,
\end{eqnarray}
where $\nu$ is the coefficient of shear viscosity. Equation (\ref{navier})
is not singular, as the coefficient of the leading 
derivative is never zero.  So in the viscous problem there is no
continuous spectrum, simply sets of discrete modes. In the zero viscosity
limit, however, the solutions to the viscous problem should reduce to the modes and
continuous spectrum of the inviscid problem. At first sight it is difficult
to see how discrete well-behaved modes could reduce to the singular
continuous spectrum solutions that we derived for the inviscid problem in
Section \ref{singsol}.  \citet{boy81} resolves this difficulty by noting
that any physical continuous spectrum perturbation is made up of an integral over frequency of
the singular inviscid solutions.  When solving the viscous
problem, the modes are discrete for even the tiniest viscosity and no integration
over frequency is required.  Solution of the viscous problem should
therefore shed light on the physical manifestation of the
continuous spectrum.

We are also pursuing the initial value problem.  It is clear from our
analysis of the normal mode problem that the time dependence of the
continuous spectrum will only be understood by solving the initial value
problem. In fact the singular continuous spectrum solutions found in this paper
only acquire physical meaning when we consider the evolution of initial
data.  We are particularly interested to see whether the 
zero-step solutions appear physically distinct in time evolutions.  That
the zero-step solutions might exhibit different time-dependence is suggested by the fact that these
solutions have zero Wronskian, which will contribute an additional pole to
the integrand of the Laplace Transform inversion integral (Watts et al, in preparation).  We also hope that
an analysis of the initial value problem will enable us to develop
necessary and sufficient conditions for instability.  

\section{Acknowledgements}

We would like to thank Luciano Rezzolla, Shin'ichirou Yoshida and Emanuele
Berti for helpful comments.  We also acknowledge support from the EU
Programme 'Improving the Human Research Potential and the Socio-Economic
Knowledge Base' (Research Training Network Contract HPRN-CT-2000-00137).
AW is supported by a PPARC postgraduate studentship and NA acknowledges
support from the Leverhulme Trust in the form of a prize fellowship and
PPARC grant PPA/G/1998/00606.

\end{document}